\documentclass[aps,twocolumn,pre,superscriptaddress]{revtex4-2}

\usepackage{dcolumn}
\usepackage{bm}

\usepackage[utf8]{inputenc}
\usepackage[T1]{fontenc}
\usepackage{mathptmx}
\usepackage{etoolbox}

\makeatother

\usepackage{xcolor,hyperref}
\hypersetup{
   colorlinks,
   linkcolor={blue!50!black},
   citecolor={blue!50!black},
   urlcolor={blue!80!black}
}

\usepackage{amsmath}
\usepackage{amssymb,changes}

\usepackage{enumerate}

\usepackage{multirow} 

\DeclareMathAlphabet{\mathitbf}{OML}{cmm}{b}{it}

\renewcommand{\=}{\!=\!}
\newcommand{\xv}{\mathitbf x}

\newcommand{\calBold}[1]{\mbox{\boldmath${\cal #1}$}}
\newcommand{\mathBold}[1]{\mbox{\boldmath$#1$}}
\newcommand{\dbar}{{\,\,\,\mathchar'26\mkern-12mu d}}

\newcommand{\Tx}{T_{\mbox{\tiny$\mathsf{X}$}}}

\definecolor{karina}{RGB}{219, 48, 122}

\usepackage{graphicx}
\begin{document}

\title{Variability of mesoscopic mechanical disorder in disordered solids}
\author{Karina Gonz\'alez-L\'opez}
\affiliation{Institute for Theoretical Physics, University of Amsterdam, Science Park 904, Amsterdam, Netherlands}
\author{Eran Bouchbinder}%
\email{eran.bouchbinder@weizmann.ac.il}
\affiliation{Chemical and Biological Physics Department, Weizmann Institute of Science, Rehovot 7610001, Israel}
\author{Edan Lerner}
\email{e.lerner@uva.nl}
\affiliation{Institute for Theoretical Physics, University of Amsterdam, Science Park 904, Amsterdam, Netherlands}

\date{\today}


\begin{abstract}
Quantifying mechanical disorder in solids, either disordered crystals or glassy solids, and understanding its range of variability are of prime importance, e.g.~for discovering structure-properties relations. The bounds on the degree of mechanical fluctuations in disordered solids and how those depend on solids formation history remain unknown. Here, we study a broadly applicable quantifier of mesoscopic mechanical disorder $\chi$, defined via the dimensionless fluctuations of the shear modulus, over a wide variety of disordered computer solids and upon varying different control parameters. $\chi$ is intimately related to basic properties of disordered solids, such as elastic constants and plastic deformability, and can be experimentally extracted by wave-attenuation measurements. We find that a large variety of \emph{self-organized} glassy solids, where disorder is an emergent property, appear to satisfy a generic lower bound on $\chi$. On the other hand, we show that $\chi$ is unbounded from above, and may diverge in systems driven towards the critical \emph{unjamming point}. These results highlight basic properties of disordered solids and set the ground for systematically quantifying mechanical disorder across different systems.
\end{abstract}

\maketitle

\section{Introduction}
\label{sec:Intro}

\vspace{-0.3cm}

Disorder has profound implications for the properties of materials. It appears in various forms and emerges from a broad range of physical processes. In the context of solids, i.e.~materials that feature a finite shear modulus, one can consider disordered crystals and glassy solids. Disordered crystals feature long-range crystalline order, but short-range disorder in the form of impurities, vacancies, interstitials and other structural and/or compositional defects. Glasses, and other amorphous solids such as granular materials, are solids in which crystallization can be entirely avoided. Such materials offer ever-growing opportunities for a huge range of technological applications and at the same time continue to pose some of the deepest scientific puzzles~\cite{debenedetti_stillinger_nature_2001,engineering_materials,wang_review_2019}. 

At the heart of both the technological applicability and basic scientific challenges resides the intrinsically non-equilibrium and disordered nature of glasses. When a glass is formed by quickly cooling a liquid below its melting point, it avoids crystallization and attains a non-equilibrium disordered structure. The latter is not unique, but rather depends on the formation process itself and consequently glasses of the very same composition can feature widely different physical properties that depend on the spontaneously emerging state of disorder~\cite{Eran_mechanical_glass_transition,experimental_inannealability,Ediger_review_2017,Ozawa6656}. Similar history dependence emerges when a packing of grains is compressed to form a granular solid~\cite{EPSTEIN1962,random_loose_packing}. Quantifying disorder in disordered solids, in particular understanding its range of variability and the corresponding variability in the emerging physical properties --- such as elastic stiffness, plastic deformability, sound attenuation coefficient and heat transport --- pose great challenges~\cite{experimental_inannealability,sticky_spheres_part_1,scattering_jcp_2021,LB_two_level_systems_prl_2020}.

\begin{figure*}
\includegraphics[width=0.935\textwidth]{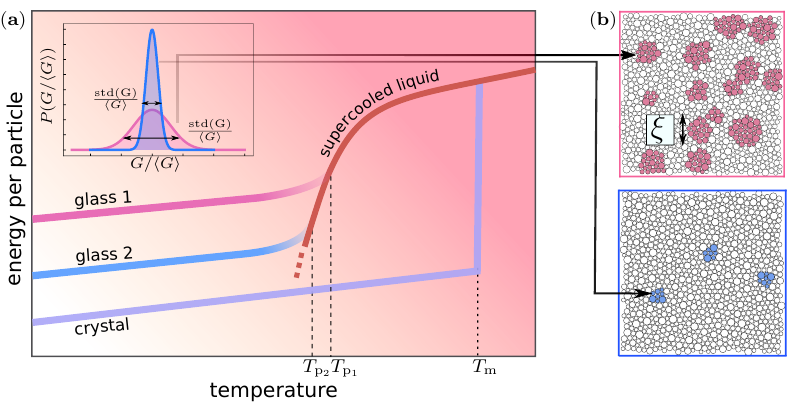}
\caption{\small{Quantifying glass preparation history, mechanical disorder and intrinsic length scale.
(a) The potential energy per particle of a glass-forming liquid is schematically plotted against its temperature. (inset) Schematic distributions of the shear modulus $G$, $P(G/\langle G\rangle)$, corresponding to the two $T_{\rm p}$ values specified on the main panel. The relative width of the distribution allows to define the dimensionless measure of mechanical disorder $\chi\!\propto\!\mbox{std}(G)/\langle G\rangle$. (b)  The upper (lower) panel corresponds to a snapshot of a glass generated using a computer glass-forming liquid model~\cite{scattering_jcp} at $T_{{\rm p}_1}$ ($T_{{\rm p}_2}$), where soft quasilocalized modes are illustrated by pink (blue) particles. See text for details.}}
\label{fig:prep_protocol}
\end{figure*}

Computer models of disordered solids offer a unique platform for addressing basic questions in materials research. One notable advantage of computer glasses is that their preparation process can be carefully controlled and quantified. In the context of glass-forming liquids, such a preparation protocol is illustrated in Fig.~\ref{fig:prep_protocol}. At high temperatures, above the melting temperature $T_{\rm m}$, the material is in an equilibrium liquid state. If the liquid is then cooled sufficiently slowly, it crystallizes at $T_{\rm m}$ (purple line). If, on the other hand, cooling is sufficiently fast, crystallization at the melting point is avoided and the supercooled liquid regime is entered (red line). At each temperature along the supercooled equilibrium line, to be denoted hereafter by $T_{\rm p}$, a glass can be formed by instantaneously cooling the liquid to a mechanically stable state at zero temperature. This process is illustrated by the two lines branching off the supercooled equilibrium line (pink line, corresponding to $T_{{\rm p}_1}$, and blue line, corresponding to $T_{{\rm p}_2}\!<\!T_{{\rm p}_1}$) in Fig.~\ref{fig:prep_protocol}a. $T_{\rm p}$ in this well-controlled procedure can be regarded as the temperature at which the glass falls out of equilibrium.

Recent developments in computer glass techniques allow to vary $T_{\rm{p}}$ over a broad range~\cite{fsp} and to follow the supercooled equilibrium line to unprecedented depth, presumably even deeper than laboratory glasses~\cite{LB_swap_prx}. Then various physical properties can be measured as a function of the glass' thermal history, quantified by $T_{\rm p}$, for a wide variety of computer glass-forming liquids. Another major advantage of computer glasses is that they allow detailed quantitative analysis of glassy disordered structures, down to the particle (atomistic) level, which is typically inaccessible to experiments.

These unique capabilities gave rise to some important recent progress. Most relevant here is the converging evidence for the existence of low-frequency glassy excitations~\cite{modes_prl_2016,ikeda_pnas,modes_prl_2018,LB_modes_2019,pinching_pnas,modes_prl_2020,QLE_jcp_2021}, whose vibrational frequencies $\omega$ follow a universal nonphononic density of states ${\cal D}(\omega)\!\sim\!\omega^4$. These universal low-frequency glassy excitations qualitatively differ from low-frequency phonons, which follow Debye's density of states, not only in their statistics, but also in their spatial properties; notably, while low-frequency phonons are spatially extended, low-frequency glassy excitations are quasilocalized, featuring a localization length $\xi$ (cf.~Fig.~\ref{fig:prep_protocol}b) and power-law decaying fields on larger scales~\cite{modes_prl_2016,atsushi_core_size_pre}. It has been further demonstrated~\cite{cge_paper,LB_modes_2019,pinching_pnas} that the number density of excitations and their typical size $\xi$ generally decrease with decreasing $T_{\rm p}$, as illustrated in Fig.~\ref{fig:prep_protocol}b. Quite remarkably, it has been very recently demonstrated that quasilocalized excitations and their nonphononic density of states ${\cal D}(\omega)\!\sim\!\omega^4$ also characterize disordered crystals~\cite{DisorderedCrystals_PRL2022}.

The intrinsic glassy length scale $\xi$ has also been shown to correspond to a crossover in the elastic response of glasses; on lengths smaller than $\xi$, elasticity is controlled by glassy disorder, while on larger lengths, continuum elasticity is recovered~\cite{barrat_point_response_2004,Ellenbroek_pre_2009,breakdown,finite_size_modes_pre_2020}. Although $\xi$ does represent a measure of glassy disorder, it is currently not directly accessible experimentally in molecular systems. We therefore focus here on {\em mesoscopic mechanical disorder} as quantified by the relative fluctuations of the shear modulus $G$, which could potentially be accessed experimentally as described further below. 

To quantify the relative fluctuations of $G$, we envision a macroscopic disordered material that is divided into volume elements, each composed of $N$ particles. For sufficiently large $N$, we expect the different volume elements to be predominantly statistically independent. While macroscopic materials cannot be generated on the computer due to computational limitations, we can generate many realizations of volume elements composed of $N$ particles each, under exactly the same preparation process, statistically representing the different volume elements of a macroscopic piece material. We then construct the realization-to-realization probability distribution function $p(G)$ and calculate $\chi\!\equiv\!\delta{G}/\langle G\rangle$. Here, $\langle\bullet\rangle$ stands for an ensemble average and
\begin{equation}
\delta{G}\equiv\sqrt{N\langle(G-\langle G\rangle)^2\rangle}=\sqrt{N}\,\mbox{std}(G)\,,
\end{equation}
where the $\sqrt{N}$ factor renders $\chi$ independent of $N$. Such distributions and the tendency of $\delta{G}/\langle G\rangle$ to decrease as $T_{\rm p}$ decreases are illustrated in the inset of Fig.~\ref{fig:prep_protocol}a.

$\chi$ as defined above provides a dimensionless measure of mechanical disorder that is closely related (under some assumptions~\cite{scattering_jcp_2021}) to Schirmacher's `disorder parameter'~\cite{Schirmacher_prl_2007,Marruzzo2013}. Additionally, $\chi$ has been very recently shown to control the attenuation of long-wavelength waves in glasses~\cite{scattering_jcp_2021} (as predicted by Heterogeneous Elasticity Theory~\cite{Marruzzo2013}) and hence to be also related to heat transport in these materials. Consequently, $\chi$ is expected to play a major role in establishing thermo-mechanical structure-properties relations in glasses and other disordered solids.

$\chi$ is intimately related to various basic properties of disordered solids, such as mesoscopic correlation lengths~\cite{sticky_spheres_part_1}, plastic deformability~\cite{loading_geometry_mrs_2021} (see also below) and wave attenuation~\cite{scattering_jcp_2021,phonon_widths2_pre_2021}, as mentioned above. The latter relation is of particular importance as it makes $\chi$ experimentally accessible. That is, it has been shown that the rate of wave attenuation $\Gamma$ in the limit of small wave numbers (long-wavelengths), $k\!\to\!0$, takes the Rayleigh scattering form $\Gamma(k)\!\sim\!\chi^2 k^4$ (in three dimensions)~\cite{scattering_jcp_2021,phonon_widths2_pre_2021}. Consequently, $\chi$ can be experimentally extracted from the pre-factor of the measured $k^4$ dependence of the wave attenuation rate $\Gamma(k)$ is the small $k$ limit. \\

What is the range of variability of the mesoscopic mechanical disorder quantifier $\chi$? How does it vary across disordered crystals featuring different forms of short-range disorder? How does it vary across glass-formers featuring different types of interactions? How does it depend on the degree of structural-mechanical frustration often present in glassy solids? These questions are addressed in this work. We start by studying simple lattice models with tunable quenched disorder. We then consider various computational protocols to vary the mechanical noise of disordered solids, ranging from reducing the internal stresses of simple computer glasses, to simulating disordered crystals with different degrees of structural frustration. We finally consider glass-forming models in which disorder is fully self-organized, and examine the effects of interaction potential and thermal annealing on the emergent mechanical disorder. 

\begin{table*}[ht!]
\caption{\label{tab:bigtable}
\footnotesize Here we provide a concise summary of the full range of the disordered solid models considered in this work, including the solid formation procedure and disorder control parameter in each case, as well as the corresponding range of variability of the quantifier of mechanical disorder $\chi$.}
\begin{center}
\begin{tabular}{|c|c|c|c|}
 \hline
Disordered solid model & Formation protocol & Control parameter & $\chi$ variability\\
    \hline
    \multirow{3}{*}{\parbox{3.5cm}{FCC lattices with stiffness disorder}}& 
    \multirow{6}{*}{\parbox{4.5cm}{\flushleft{Unit masses placed on a lattice and interact with nearest neighbors via Hookean springs, whose stiffnesses $k$ are drawn from {(\it{i})}~Gaussian, {(\it{ii})}~Log-normal, {(\it{iii})}~Gamma distributions (see also Table~\ref{tab:distributions}).}}} &
    \multirow{7}{*}{\parbox{4cm}{Ratio $\Delta k/\bar{k}$ of the standard deviation and mean of the stiffness distribution}}&
    \multirow{2}{*}{{(\it{i})}\,\,[0,\,1.5e-3]}\\
    &&&\multirow{2}{*}{{(\it{ii})}\,\,[0,\,7.5e-3]}\\ 
    &&&\multirow{2}{*}{{(\it{iii})}\,\,[0, 2.6e-2]}\\[15pt] \cline{4-4} \cline{1-1} 
    \multirow{3}{*}{\parbox{3.5cm}{Cubic lattices with stiffness disorder}}& 
    &\multirow{3}{*}{}&\multirow{2}{*}{{(\it{i})}\,\,[0, 1.5e-3]}\\ 
    &&&\multirow{2}{*}{{(\it{ii})}\,\,[0, 6.4e-3]}\\
    &&&\multirow{2}{*}{{(\it{iii})}\,\,[0, 1.7e-2]}\\[15pt]\cline{1-4} %
    \parbox{3.6cm}{Disordered spring networks (positional \& topological disorder)}&
    \parbox{4.5cm}{\flushleft{Disordered networks are derived from the interaction-networks of soft-sphere packings. Unit masses are placed on the nodes, and edges are replaced by Hookean springs of unit stiffness. Springs are then randomly removed to reach a target connectivity $z$.}}&
    \parbox{4cm}{Excess connectivity $z\!-\!z_{\rm c}$\\ ($z_{\rm c}\equiv 2\times$ the dimension of space)}&
    [3.1e-1, 21.1]\\[50pt]
    \hline   

    \parbox{3.6cm}{Disordered crystals}&
    \parbox{4.5cm}{\flushleft{Single-component FCC lattices in which $N/2$ randomly selected particles are replaced by `impurity' particles.}}&
    \parbox{4cm}{Degree of impurity interaction mismatch $\delta$}&
    [9.5e-5, 9.7]\\[30pt]
    \hline
    \parbox{3.6cm}{Fluctuating size particles} &
    \parbox{4.5cm}{\flushleft{During glass formation alone, particles’ effective sizes are dynamic degrees of freedom subjected to a potential of characteristic stiffness $k_\lambda$.}}&
   \parbox{4cm}{Stiffness associated with the effective particle size $k_{\lambda}$}&
    \,\,[8.9e-1, 3.6]\\[35pt]
    \hline
   \parbox{3.6cm}{Internal-stress-reduced glasses}&
    \parbox{4.5cm}{\flushleft{Frustration-induced internal stresses are down-scaled in simple soft-sphere glasses.}}&
    \parbox{4cm}{Fractional degree of internal-stress reduction $\Delta$}
    &\,\,[4.01e-1, 4.03]\\[25pt]
    \hline
    \multirow{2}{*}{\parbox{3.5cm}{Sticky hard-spheres glasses}}&
    \parbox{4.5cm}{\flushleft{Glasses formed by instantaneously quenching high-temperature liquids.}}&
    \parbox{4cm}{Short-range repulsion parameter $Q$}&
    \multirow{3}{*}{[3.6, 6.3]}\\[18pt]
    \hline
    \multirow{3}{*}{\parbox{3.6cm}{Kob-Andersen Binary Lennard-Jones}}&
    \multirow{1}{*}{\parbox{4.5cm}{\flushleft{Glasses formed by instantaneously quenching liquids equilibrated at various parent temperatures $T_{\rm p}$, using Molecular Dynamics.}}}& 
    \multirow{13}{*}{\parbox{4cm}{Properly normalized parent temperature (glass transition temperature) $T_{\rm p}/\Tx$}} &
    \multirow{3}{*}{[2.8, 4]}\\[20pt]\cline{1-1}\cline{4-4}
    
    \multirow{3}{*}{\parbox{3.6cm}{Sticky spheres (binary) 1}}&&&\multirow{3}{*}{[1.4, 1.8]}\\[15pt]\cline{1-1}\cline{4-4}
    \multirow{3}{*}{\parbox{3.6cm}{Sticky spheres (polydisperse) 1}}&&&\multirow{3}{*}{[9.2e-1, 1.2]}\\[20pt]\cline{1-1}\cline{2-2}\cline{4-4}
    \parbox{3.6cm}{Polydisperse soft spheres} &
    \parbox{4.5cm}{\flushleft{Glasses formed by instantaneously quenching liquids equilibrated over a very large range of parent temperatures $T_{\rm p}$, using the Swap-Monte-Carlo algorithm~\cite{LB_swap_prx}.}}& 
    &
    \,\,[9.3e-1, 3.7]\\[35pt] 
    \hline
\end{tabular}
\end{center}
\label{tab:multicol}
\end{table*}

This work is structured as follows; in Sect.~\ref{sec:chi_calculation}, we describe the difference between `gapped' and `gapless' disordered solids, and how the mechanical disorder quantifier $\chi$ is calculated for each of these classes of systems. In Sect.~\ref{sec:results}, we describe our comprehensive study of mechanical noise in a broad variety of models of disordered solids, from the simplest models of quenched disorder, to the fully self-organized structural glass models. The full range of disordered solids considered and our main results are concisely summarized in Table~\ref{tab:bigtable}. We discuss our results and provide an outlook for future work in Sect.~\ref{sec:results_and_discussion} . Details about the numerical models and protocols are provided in the Appendices.

\section{Calculating the quantifier of mechanical disorder $\chi$}
\label{sec:chi_calculation}

In this work, we focus on the fluctuations of the shear modulus $G$. In the athermal limit, $T\!=\!0$, it reads~\cite{lutsko}
\begin{equation}\label{eq-G}
    G \equiv \frac{1}{V}\frac{d^2U}{d\gamma^2}= \frac{1}{V}\left(\frac{\partial^{2}U}{\partial \gamma^{2}}-\frac{\partial^{2}U}{\partial \gamma\partial \xv} \cdot \calBold{M}^{-1}\cdot \frac{\partial ^{2}U}{\partial\xv\partial \gamma}\right)\,,
\end{equation}
where $\xv$ denotes particles' coordinates, $\calBold{ M}\!\equiv\!\frac{\partial^2U}{\partial\xv\partial\xv}$ denotes the Hessian matrix of the potential $U(\xv)$, and $\gamma$ is a shear-strain parameter that parameterizes the imposed affine simple shear (in the $x$-$y$ plane) transformation of coordinates $\xv\!\to \mathBold{H}(\gamma)\cdot\xv$ with
\begin{equation}\label{shear_transformation_matrix}
\mathBold{H}(\gamma) =  \left( \begin{array}{ccc}1&\gamma&0\\0&1&0\\
0&0&1\end{array}\right)\,.
\end{equation}

When calculating the mechanical disorder quantifier $\chi$ for different classes of solids, we make a distinction between gapped and pseudo-gapped systems; we refer to systems as being \emph{gapped} if the frequencies of \emph{nonphononic} vibrational modes are bounded from below, as occurs e.g.~in systems of relaxed Hookean-springs~\cite{mw_EM_epl}. In contrast to gapped systems, in \emph{pseudo-gapped} systems nonphononic vibrational modes can occur at any frequency --- no matter how small ---, as their density of states grows from zero frequency as $\sim\!\omega^4$, and see~\cite{QLE_jcp_2021} for a recent review of pseudo-gapped glasses that host low-frequency quasilocalized vibrational modes.

These aforementioned spectral differences lead to different realization-to-realization distributions of the shear modulus $p(G)$; in particular, earlier work~\cite{scattering_jcp_2021} has shown that $p(G)\!\sim\!(\bar{G}\!-\!G)^{-7/2}$ in structural glasses (pseudo-gapped systems), with an $N$-dependent prefactor that vanishes as $\sim\!N^{-3/2}$ (here $\bar{G}$ denotes the ensemble-mean shear modulus). In finite volume realizations, these anomalous statistics can hinder a proper measurement of the width of $p(G)$, as described at length in Ref.~\cite{sticky_spheres_part_1}. 

To overcome this difficulty, we follow~\cite{sticky_spheres_part_1,scattering_jcp_2021} and use a Jackknife-like method that consists of calculating $\chi$ via:
\begin{equation}
\label{eq:chi}
    \chi = \frac{\sqrt{N\langle \left( G-\langle G\rangle\right)^{2}\rangle}}{\langle G\rangle} \ ,
\end{equation}
using the entire ensemble of glasses at hand, and here $\langle \bullet\rangle$ represents an ensemble-average. Then, for each data point $G_{i}$, we calculate $\chi^{i}$ using Eq.~\eqref{eq:chi}, but excluding the $i^{\rm th}$ data point $G_{i}$ from the calculation. We then compare the percent difference $\delta_{i}\!\equiv\!100\!\times\!(\chi\!-\!\chi^{i})/\chi$ for each data point $i$, permanently remove the data point with the {\it{largest}} percent difference $\delta_{i}$ from the total data set, and re-calculate $\chi$. This algorithm is repeated until the maximal percent difference is less than $\delta\!=\!1\%$. This scheme, as well as the chosen percentage for $\delta$, are motivated and explained in more detail in~\cite{sticky_spheres_part_1}. Finite-size effects in $\chi$ for pseudo-gapped systems are discussed in Appendix~\ref{sec:appendix_finitesize_chi}.

Finally, in gapped systems, we calculate $\chi$ following the definition provided in Eq.~(\ref{eq:chi}), without any exclusion or post-processing.

\section{Mechanical noise in disordered solids}
\label{sec:results}

In this Section, we report on a comprehensive analysis of the mechanical disorder --- as quantified by $\chi$ --- in a rather broad range of computer models of disordered solids. We start from relatively simple lattice models with quenched disorder (stiffness, positional and topological), and move on to consider glassy solids generated by various formation protocols, culminating with a study of models of structural glasses in which disorder is spontaneously emergent. Our results are first concisely summarized in Table~\ref{tab:bigtable}, where the full range of the disordered solid models we considered is presented. We also briefly describe therein the solid formation procedure and disorder control parameter in each case, as well as the corresponding range of variability of the quantifier of mechanical disorder $\chi$. The results themselves are described and discussed in the subsequent subsections. 


\begin{figure}[hb]
\includegraphics[width=1.0\linewidth]{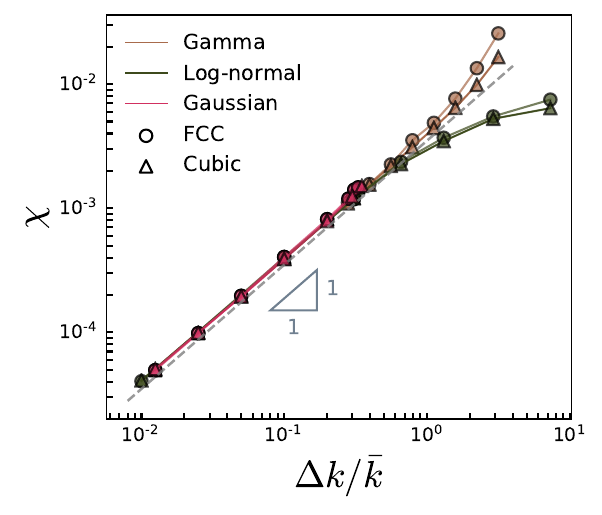}
\caption{\small{Mechanical disorder quantifier $\chi$~\emph{vs.}~dimensionless ratio $\Delta k/\bar{k}$ of the three spring stiffnesses' distributions (see Table~\ref{tab:distributions}) for both FCC (circles) and cubic (triangles) lattices. }}
\label{fig:chi_spring_lattices}
\end{figure}

\vspace{5cm}

\subsection{FCC and cubic spring lattices with stiffness disorder}\label{sec:lattices}

We launch the discussion by considering FCC and cubic lattices (with the short diagonals included) of unit masses connected by relaxed Hookean springs whose stiffnesses $k$ are drawn from a distribution $p(k)$ characterized here by its width (standard deviation) to mean ratio $\Delta k/\bar{k}$. We employ three different distributions for $p(k)$: 

\begin{itemize}
\item[(\it{i})] Gaussian: $p(k) = \frac{1}{\sqrt{2\pi\sigma^2}}e^{-(k-k_0)^2/2\sigma^2}$
\item[(\it{ii})] Log-normal: $p(k) = \frac{1}{k\sqrt{2\pi\sigma^2}}e^{-(\ln k-k_0)^2/2\sigma^2}$
\item[(\it{iii})] Gamma: $p(k) = \frac{1}{\Gamma(\theta)k_0}(k/k_0)^{\theta-1}e^{-k/k_0}$
\end{itemize}
with means, widths and width-to-mean ratios expressed in terms of the parameters $k_0,\theta$ and $\sigma$ provided in table~\ref{tab:distributions}, and $\Gamma(\cdot)$ is the Gamma function. 

\begin{table}[h!]
\centering
\caption{\label{tab:distributions}
\footnotesize Mean, width, and characteristic parameter $\Delta k/\bar{k}$ for the three spring stiffness distributions used. Note that $k_0$ set to unity here and throughout this Subsection.}
{\renewcommand{\arraystretch}{1.8}
\vspace{0.2cm}
\begin{tabular}{|c|c|c|c|}
\hline
Distribution & mean($\bar{k}$) & width($\Delta k$) & ratio($\Delta k/\bar{k}$)\\
\hline
Gaussian & $k_0$ &  $\sigma$ & $\sigma/k_0$\\
\hline
Log-normal & $e^{k_0+\sigma^2/2}$ & $e^{k_0+\sigma^2/2}\sqrt{e^{\sigma^2}\!-\!1}$ & $\sqrt{e^{\sigma^2}\!-\!1}$\\
\hline
Gamma & $k_0\theta$ & $k_0\sqrt{\theta}$ & $1/\sqrt{\theta}$\\
\hline
\end{tabular}
}
\end{table}

In Fig.~\ref{fig:chi_spring_lattices}, we plot $\chi$ vs.~the dimensionless parameter $\Delta k/\bar{k}$ ($\rm{std}(k)$-to-$\rm{mean}(k)$ ratio) for the six systems studied; as predicted in~\cite{phonon_widths,phonon_widths2_pre_2021} based on a perturbative approach, we find that $\chi\sim \Delta k/\bar{k}$. This scaling persists up to $\Delta k/\bar{k}\!\approx\!0.6$, for all distributions and both lattice types, which apparently marks the onset of a strong disorder regime, which cannot be described perturbatively. Interestingly, the prefactor of the $\chi\sim \Delta k/\bar{k}$ scaling law appears to be very similar for both FCC and cubic lattice types. 

\subsection{Spring networks with positional and topological disorder}\label{sec:networks}

We next consider disordered networks of relaxed Hookean springs connecting (unit) point masses, with both positional and topological (i.e.~degree of connectivity) disorder. This is achieved by adopting the interaction networks of simple, three-dimensional (3D) soft-spheres glasses (see Ref.~\cite{cge_paper} for a description of the soft-spheres model), where we place a Hookean spring between every pair of interacting particles in the original glass. This procedure results in a disordered spring network of initial coordination $z\!\approx\!16$, which is much larger than the Maxwell threshold $z_{\rm c}\=6$ in 3D. We then systematically remove bonds by considering in each iteration the bond $i,j$ whose combined connectivity $z_i\!+\!z_j$ is largest. Since there are many bonds that share the same combined connectivity $z_i\!+\!z_j$, we consider a secondary bond-removal criterion: amongst all bonds $i,j$ whose $z_i\!+\!z_j$ is maximal, we select to remove a bond whose difference $|z_i\!-\!z_j|$ is smallest. These two criteria ensure that the connectivity fluctuations of the resulting disordered spring network are small. 

\begin{figure}[b]
\includegraphics[width=1.0\linewidth]{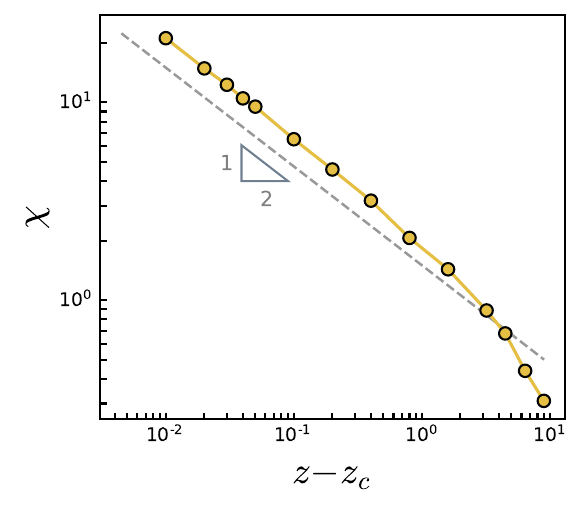}
\caption{\small{Mechanical disorder $\chi$ of random Hookean-spring networks of $N\!=\!131\,072$ nodes in three dimensions ($\dbar\!=\!3$), plotted as a function of the distance $z\!-\!z_{\rm c}$ to the critical connectivity $z_{\rm c}\!\equiv\!2\,\dbar$ at which unjamming occurs~\cite{liu2011jamming,van_hecke_review}. Mechanical disorder diverges when the unjamming transition is approached and the expected scaling~\cite{phonon_widths2_pre_2021} $\chi\!\sim\!(z\!-\!z_{\rm c})^{-1/2}$ is validated.}}
\label{fig:chi_deltaz_networks}
\end{figure}

Disordered networks of relaxed Hookean springs are known to undergo an `unjamming transition'~\cite{liu2011jamming,van_hecke_review} when the mean connectivity $z$ approaches the Maxwell threshold~\cite{maxwell_1864} $z_{\rm c}\=2\,\dbar$, where $\dbar$ is the spatial dimension. In Fig.~\ref{fig:chi_deltaz_networks}, we show that mechanical disorder as quantified by $\chi$ diverges in the limit $z\!\to\!z_{\rm c}$ as $\chi\!\sim\!(z\!-\!z_{\rm c})^{-1/2}$, in agreement with arguments and simulations in two-dimensions of Ref.~\cite{phonon_widths2_pre_2021}, and with previous results of Ref.~\cite{Goodrich_pre_2014}. We thus establish that tuning mechanical disorder to be arbitrary large can be achieved by approaching the unjamming point.

\subsection{Crystals with compositional disorder (disordered crystals)}\label{sec:disordered_crystals}

We next turn to examining the degree of mechanical disorder in a computer model of disordered crystals. In this model, some fraction $x$ (here, we consider $x\!=\!1/2$) of the particles of a perfect, single-species FCC lattice are replaced with particles of a second species (`impurity' particles), following Barrat et al.~\cite{mizuno2014acoustic,mizuno2016relation}. For these simulations, we employ a simple system of soft spheres interacting via an $r^{-10}$ pairwise potential, as used e.g.~in Ref.~\cite{cge_paper}. We start the procedure by placing $N\!=\!131072$ mono-dispersed particles on a FCC lattice, and selecting randomly a set of $N/2$ particles that will be replaced with the larger-particle species in order to induce an amorphization transition. At every step, we incrementally vary the parameter $\delta\!\in\![0,1]$ that determines the differences between the two particle species in the system; $\delta\!=\!0$ corresponds to the untouched, single-species crystal, while $\delta\!=\!1$ recovers the 1:1.4 `small'-`large' size ratio used in many computer models of liquids and glasses, e.g.~the model of Ref.~\cite{cge_paper}. Effectively, $\delta$ quantifies the  interaction mismatch with impurity particles. Further details are provided in Appendix~\ref{sec:appendix_disordered_crystals}. 

\begin{figure}[h!]
\includegraphics[width=1.0\linewidth]{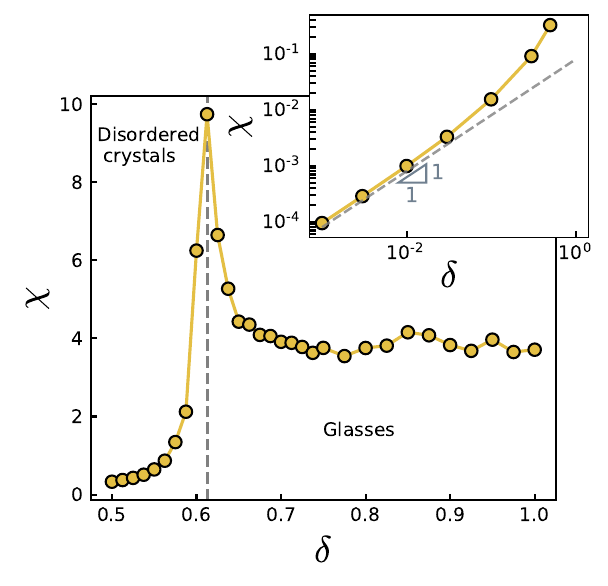}
\caption{\small{The mechanical disorder quantifier $\chi$ is plotted against the dimensionless parameter $\delta$ that controls the amorphization transition in a model of disordered crystals; $\delta\!=\!0$ correponds to no disorder, while $\delta\!=\!1$ recovers a binary mixture of soft spheres with size ratio 1:1.4. Inset: the scaling of $\chi\!\sim\!\delta$ is seen at small $\delta$.}}
\label{fig:chi_delta_amorphization}
\end{figure}

In Fig.~\ref{fig:chi_delta_amorphization}, we report $\chi$ vs.~$\delta$, where averages were taken over 1000 independent disordered crystals for each~$\delta$. We observe that $\chi(\delta)$ is non monotonic, with a sharp peak marking the amorphization transition located at $\delta\!=\!0.6125$. Furthermore, we find that $\chi$ grows linearly with the amorphization parameter $\delta$ for very small $\delta$ values (see inset of Fig.~\ref{fig:chi_delta_amorphization}). Related results concerning the degree of mechanical disorder in disordered crystals, and the relations between the latter and structural glasses, can be found in~\cite{DisorderedCrystals_PRL2022}.

\subsection{Internal-stress-reduced glasses}\label{sec:one_minus_delta}

Internal stresses and structural frustration in glasses have been shown to give rise to soft excitations~\cite{eric_boson_peak_emt,inst_note}, enhanced responses to local perturbations~\cite{breakdown}, and mesoscale mechanical fluctuations~\cite{scattering_jcp_2021}. Here, we consider initial glassy states that are created by quenching a liquid, where internal stresses are subsequently varied in a controlled manner. This allows us to study the effects of internal stresses in a generic glass-forming model on the range of variability of $\chi$, as explained below. This computational approach was also used to establish a fundamental relation between $\chi$ and long-wavelength wave-attenuation rates~\cite{scattering_jcp_2021}, in addition to in other contexts~\cite{breakdown,eric_boson_peak_emt,inst_note}.

\begin{figure}[ht!]
\includegraphics[width=0.95\linewidth]{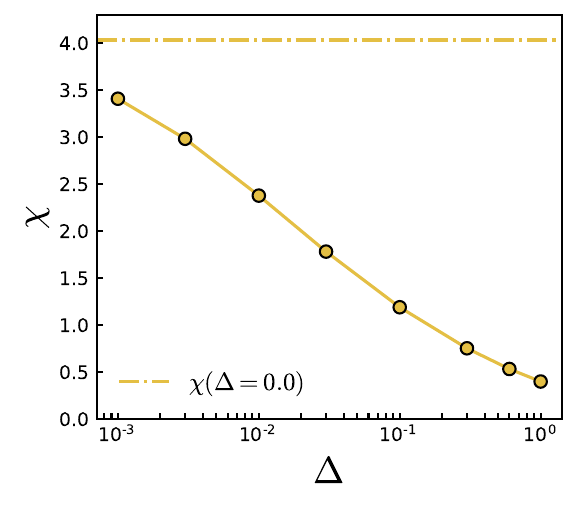}
\caption{\small{Mechanical disorder quantifier $\chi$ for a glass-forming model of soft-spheres whose internal stresses have been controllably reduced by increasing the parameter $\Delta$, see main text for precise definition. The dash dotted line marks $\chi$ of the original glasses, before reducing the internal stresses (corresponding to $\Delta\!=\!0$).}}
\label{fig:chi_vs_delta_one_minus_delta}
\end{figure}

To study the variability of $\chi$ with variations in the internal stresses of glasses, we employ the same simple soft-spheres glasses used to create the initial networks in Sect.~\ref{sec:networks}, see e.g.~Ref.~\cite{cge_paper}. This model features $\chi\!\approx\!4$, before any reduction of internal stresses has been carried out. To proceed, in each glass sample we rescale the interparticle forces by a factor $1\!-\!\Delta$ ($0\!\le\!\Delta\!\le\!1$), and recalculate the shear moduli of the solids, as done previously in Refs.~\cite{breakdown,eric_boson_peak_emt,inst_note,scattering_jcp_2021}. In Fig.~\ref{fig:chi_vs_delta_one_minus_delta}, we plot $\chi(\Delta)$ and observe that as $\Delta$ is increased and internal stresses are reduced, mesoscopic mechanical disorder decreases, reaching $\chi\!\approx\!0.4$ in the systems with no internal stresses ($\Delta\!=\!1$). 

\begin{figure*}
\includegraphics[width=1.0\linewidth]{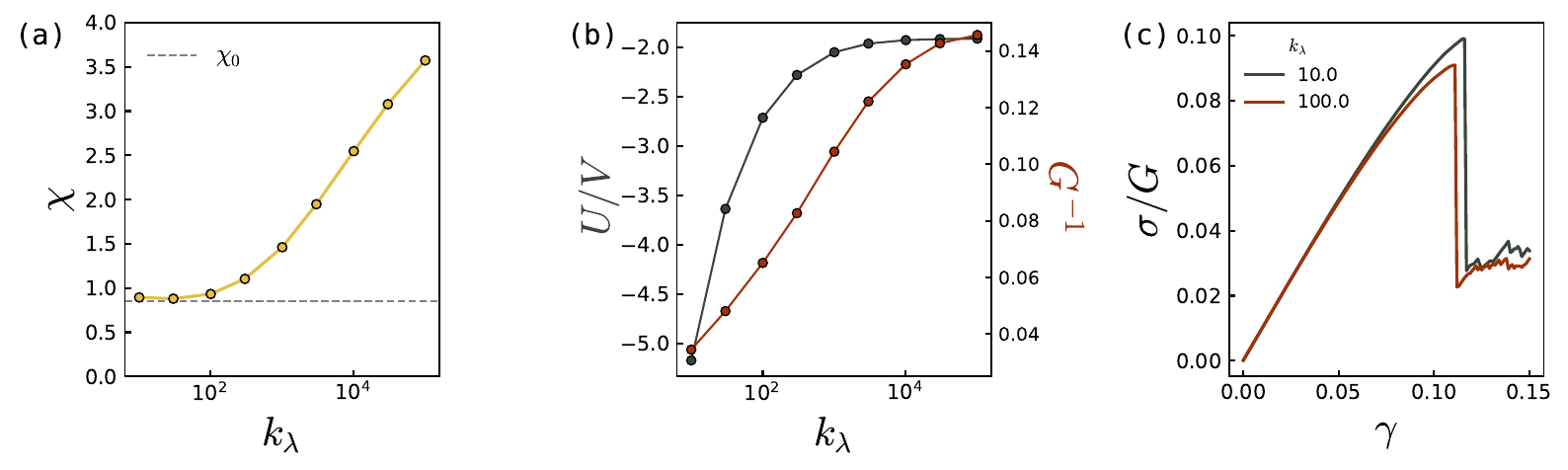}
\caption{\small{(a) The mechanical disorder quantifier is plotted for FSP glasses~\cite{fsp}, vs.~the parameter $k_\lambda$ that represents the stiffness associated with the fluctuating effective sizes of particles during glass formation. (b) Potential energy density $U/V$ (left $y$-axis, green data points) and shear compliance $ G^{-1}$. (c) The stress normalized by the average shear modulus $\sigma/G$ \emph{vs.~}strain $\gamma$ curves for the two most stable glasses employed; while they both share very similar values of $\chi$, their nonlinear macroscopic response is measurably different, see text for further discussion.}}
\label{fig:chi_fsp}
\end{figure*}

\vspace{5cm}
\subsection{Fluctuating-size-particles glasses} \label{sec:fsp}

We now turn to study the variability of mechanical disorder in glasses that are \emph{not} formed by quenching a liquid --- but nevertheless still feature \emph{self-organized} structures. One interesting instance of such a system is a computer glass former~\cite{fsp} in which the particles’ effective sizes are considered to be degrees of freedom subjected to an internal potential that favors some particular size. The glasses created with this model are formed by a rapid quench from high energy states, after which their effective-size degrees of freedom are frozen out. A tunable parameter, $k_\lambda$, controls the stiffness associated with the effective-size degrees of freedom, and plays a crucial role in determining the stability of the resulting computer glasses. In particular, for $k_\lambda\!\to\!\infty$ the effective-size degrees of freedom are entirely frozen, and a generic computer glass is recovered, while for smaller $k_\lambda$ the effective sizes allow for additional relaxation and stabilization. We employed the same glass ensembles studied in Ref.~\cite{fsp}; technical details can be found there and in Appendix~\ref{sec:appendix_models}. We refer to this model as the Fluctuating-Size Particles (FSP) model.

We extract the mechanical disorder quantifier $\chi$ under variations of $k_\lambda$ by several orders of magnitude.  Our results are summarized in Fig.~\ref{fig:chi_fsp}; we observe a dramatic decrease of $\chi$ as the stability of the computer glasses increases (by decreasing $k_\lambda$). Strikingly, the mesoscopic mechanical disorder appears to be bounded from below (see Fig.\ref{fig:chi_fsp}a). Nevertheless, at the same time other observables continue to evolve with $k_\lambda$ after $\chi$ has saturated. This is illustrated by the energy density $U/V$ (green points) and the shear compliance $G^{-1}$ (terracota points) plotted in Fig.~\ref{fig:chi_fsp}b, which continue to decrease even for the $k_\lambda$ values where $\chi$ has reached a plateau. Finally, in Fig.~\ref{fig:chi_fsp}c, we show the stress-strain curves for glasses of $N\!=\!125,000$ particles using $k_\lambda\!=\!10$ and $k_\lambda\!=\!100$ to emphasize that these two glasses show different mechanical response. We find, as expected, the less stable glass $k_\lambda\!=\!100.0$ to feature a smaller stress drop, despite that $\chi(k_{\lambda}\!=\!10.0)\!\approx\!\chi(k_{\lambda}\!=\!100.0)$.


\begin{figure}[hb]
\includegraphics[width=1.0\linewidth]{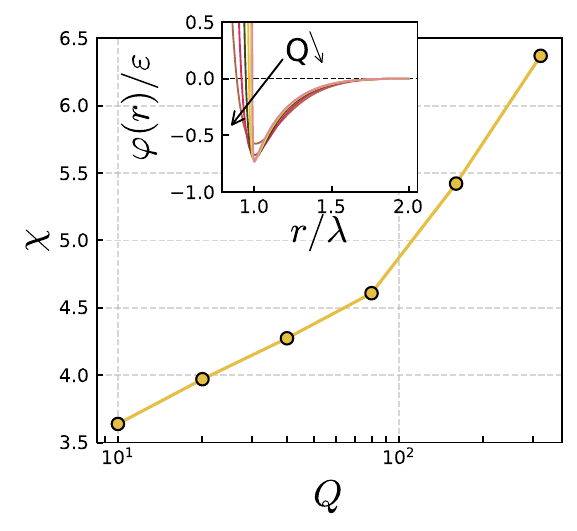}
\caption{\small{The mechanical disorder quantifier $\chi$ is plotted vs.~the parameter $Q$ that determines the steepness of the pairwise potential of the Sticky-Hard-Spheres model, as demonstrated in the inset. Taking $Q\!\to\!\infty$ corresponds to approaching the unjamming point (as explained at length in~\cite{ultrahigh_poissons_ratio_prm2022}), and therefore $\chi$ grows with $Q$.}}
\label{fig:chi_q_stickyhs}
\end{figure}

\subsection{Glasses quenched from a melt: Sticky hard spheres with variable short-range repulsion}\label{sec:sticky_hs}

We next move ahead to consider systems in  which disorder is spontaneously self-generated during the glass formation process, rather than being  externally controlled as in the previous subsections. We start by presenting the results from a simple glass-forming model of a binary mixture in 3D, whose particles interact via a pairwise potential in which the stiffness associated with the short-range repulsion can be tuned via the parameter $Q$. As shown in the inset of Fig.~\ref{fig:chi_q_stickyhs}, decreasing $Q$ leads to a softer repulsive part of this potential. Further details about the model are provided in Appendix~\ref{sec:appendix_models}. This system is particularly interesting since it was recently shown~\cite{ultrahigh_poissons_ratio_prm2022} that, via an unjamming-like mechanism, glasses with large $Q$ feature ultra-high Poisson's ratio, exceeding values of any laboratory metallic glass reported to date. Details about the glass-formation protocol and system sizes can be found in Ref.~\cite{ultrahigh_poissons_ratio_prm2022} and Appendix~\ref{sec:appendix_models}.

The main panel of Fig.~\ref{fig:chi_q_stickyhs} shows the variability of the realization-to-realization fluctuations of the shear modulus --- quantified by $\chi$ --- as as function of $Q$. Our data show that $\chi$ varies by a factor of approximately $2$ for the studied range of $Q$'s. Similarly to the disordered spring networks, also here $\chi$ grows as the unjamming point $Q\!\to\!\infty$ is approached. Further discussions about the properties of this model that approaches the sticky-hard-spheres limit can be found in~\cite{ultrahigh_poissons_ratio_prm2022}. 

\subsection{Glasses quenched from a melt: Various structural glass model with variable thermal history}\label{sec:melts}

\begin{figure}
\includegraphics[width=1.1\linewidth]{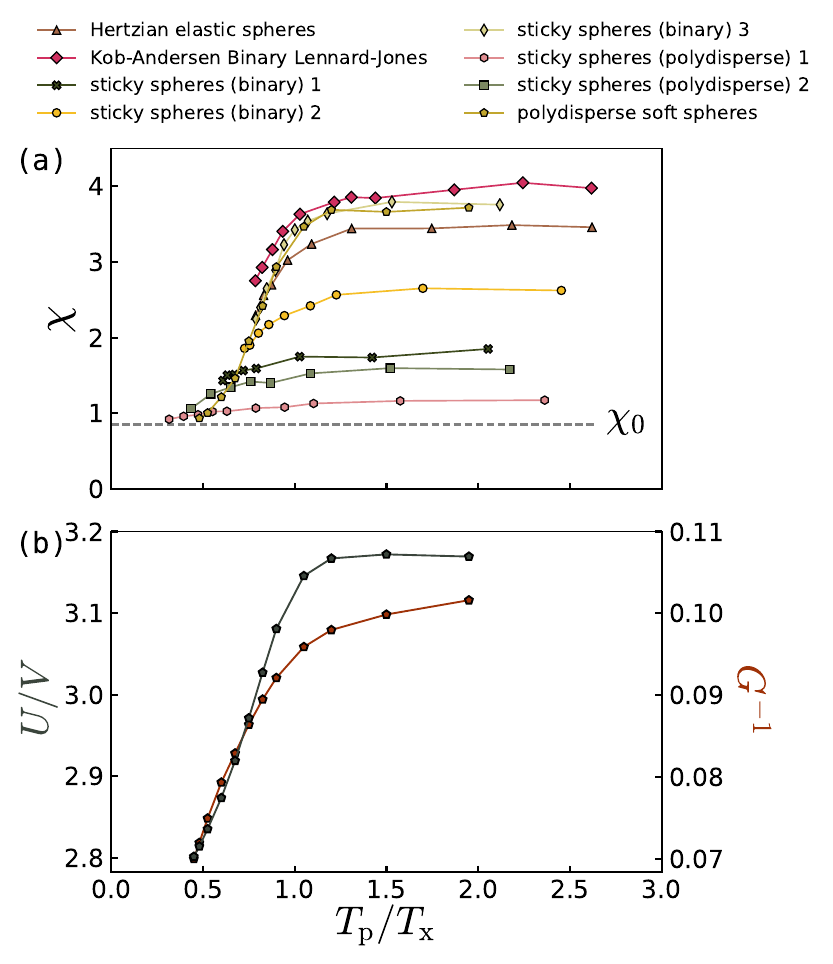}
\caption{\small{(a) $\chi(T_{\rm p}/\Tx)$ curves of the different models converge to a master envelope curve, which appears to be bounded from below. (b) The potential energy density $U/V$ (left $y$-axis, green data points) and the shear compliance $G^{-1}$ for the polydisperse soft spheres model, as a function of $T_{\rm p}/\Tx$} over the same temperature range as in (a).}
\label{fig:chi_tptx}
\end{figure}

Here, we investigate variability of mechanical disorder in structural glasses quenched from a melt. We consider ensembles of glasses labeled by the parent equilibrium temperature $T_{\rm p}$ from which those glasses were instantaneously quenched. Consequently, $T_{\rm p}$ plays the role of an effective glass transition temperature. We employ several different computer glass-forming  models (listed in the legend of Fig.~\ref{fig:chi_tptx}) featuring a wide variety of interparticle potentials --- from purely repulsive to strongly attractive --- and different compositions --- from binary mixtures to broadly polydispersed. In Appendix~\ref{sec:appendix_models}, we provide additional details about the pairwise potentials, system sizes, number densities, and number of independent glass samples that constitute each ensemble.

In Fig.~\ref{fig:chi_tptx}, we present our results; in order to compare the $T_{\rm p}$-dependence of $\chi$ between different models on the same footing, for each model we estimated a crossover temperature scale $\Tx$ (following the procedure put forward in Ref.~\cite{Tx_jcp}), and plot $\chi$ vs.~the rescaled parent temperature $T_{\rm p}/\Tx$. $T_{\rm{p}}$ was  varied  from  high  temperature  liquid states to deeply supercooled states, the latter correspond to the limit of our computational capacity (using Molecular Dynamics for binary mixtures, or Swap Monte Carlo~\cite{LB_swap_prx} for polydisperse systems). We find that high-$T_{\rm{p}}$ $\chi$ values vary substantially among the different models, confirming their wide variability. Additionally, we find that data from all models follow a master envelope, lending further support to the effectiveness of $\Tx$ in comparing elastic properties among different glasses quenched from a melt. 

Our results also show that mechanical noise decreases as the stability of glasses increases, with the variability of $\chi$ in each model seemingly dependent on the high-$T_{\rm p}$ value. Interestingly, our data appear to suggest that $\chi$ might feature a \emph{lower bound}, putatively marked by the dashed horizontal line in Fig.~\ref{fig:chi_tptx}a. To highlight this observation, we plot in Fig.~\ref{fig:chi_tptx}b the energy density $U\!/\!V$ (left axis, green points) and the shear compliance $G^{-1}$ (right axis, brown points) of the polydisperse soft spheres (that can be deeply supercooled using Swap-Monte-Carlo~\cite{LB_swap_prx}) vs.~the rescaled parent temperature $T_{\rm p}/\Tx$; while for this system $\chi$ appears to decrease with a noticeable positive curvature (see pentagonal symbols in Fig.~\ref{fig:chi_tptx}a), the energy density and shear compliance show no measurable signs of saturation.

\begin{figure}[ht!]
\includegraphics[width=1.0\linewidth]{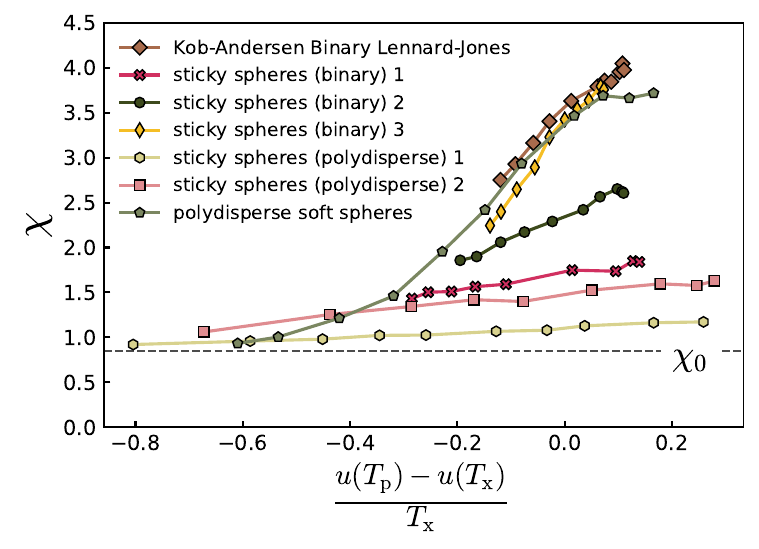}
\caption{\small{Mechanical disorder $\chi$ is plotted vs.~the shifted and rescaled potential-energy-per-particle $(u(T_{\rm p})\!-\!u(\Tx))/\Tx$ for a variety of different computer glass-formers, where $\Tx$ is the crossover temperature -- defined in Ref.~\cite{Tx_jcp} and employed in this work. In this representation it becomes clear that at lower energy states the mechanical disorder as quantified by $\chi$ appears to saturate above a bound $\chi_0$ putatively marked by the dashed horizontal line.}}
\label{fig:chi_utx}
\end{figure}

\begin{figure}
\includegraphics[width=1.0\linewidth]{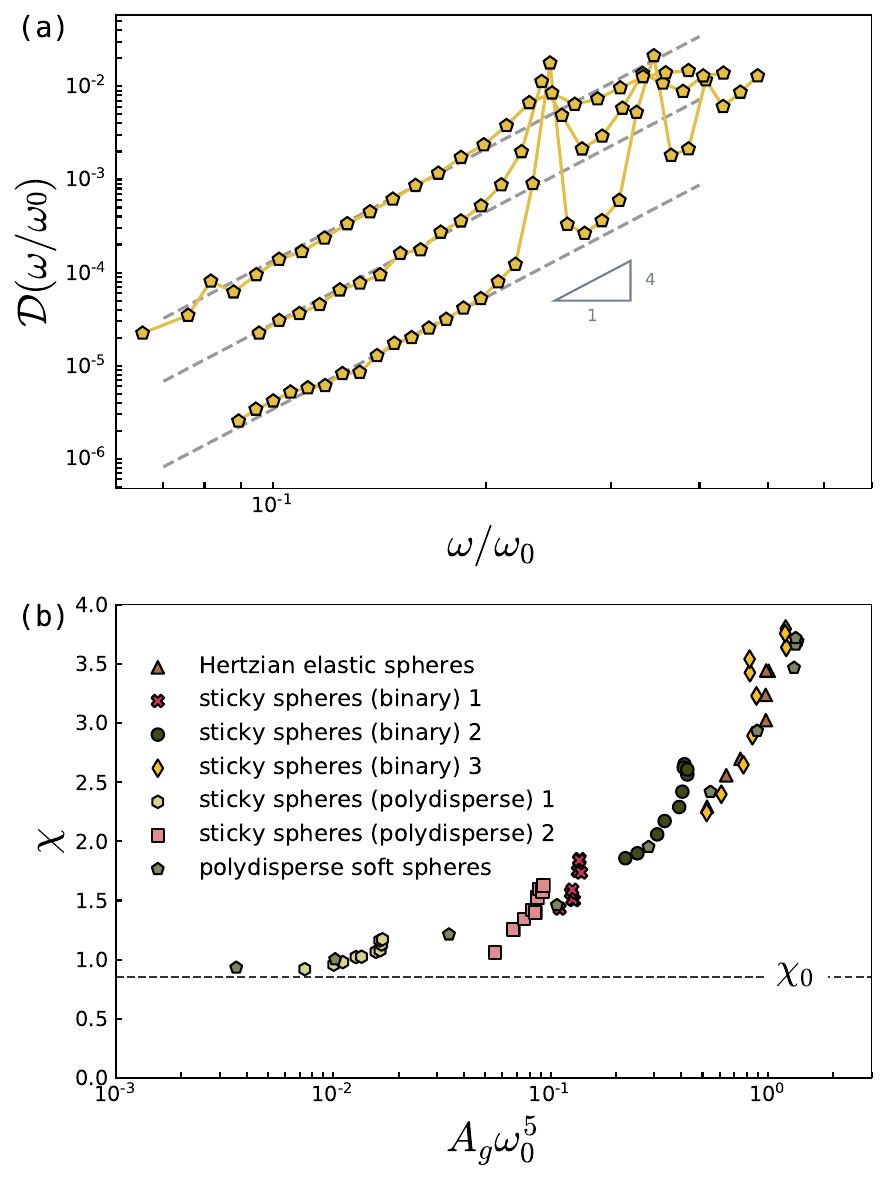}
\caption{\small{(a) The prefactor $A_{\rm g}$ of the universal ${\cal D}(\omega)\!=\!A_{\rm g}\omega^4$ nonphononic spectrum is extracted as demonstrated in the figure for $T_{\rm p}/\Tx\!=\!1.95,0.75,0.60$ of the `polydisperse soft spheres' model. (b) $\chi$ is plotted against $A_{\rm g}$, nondimensionalized by $\omega_0^5\!\equiv\!(c_s/a_0)^5$, forming an approximate master curve that appears to saturate as $A_{\rm g}\!\to\!0$.}}
\label{fig:chi_ag}
\end{figure}

It is appealing to consider the variation of $\chi$ with $T_{\rm p}$, \emph{i.e.}~$\chi(T_{\rm p})$, since $T_{\rm p}$ is in principle experimentally controlled, \emph{e.g.}~by varying the glass formation cooling rate (as done in Fig.~\ref{fig:chi_tptx}). However, the super-exponential dependence of relaxation times on temperature --- characteristic of supercooled liquids~\cite{debenedetti_stillinger_nature_2001} --- implies that a very small reduction in $T_{\rm p}$ requires a very large reduction in the cooling rate. Yet, the potential energy in deeply supercooled liquids decreases rapidly as temperature is decreased. It is therefore expected that a clearer picture of the saturation of $\chi$ at a lower bound might emerge when plotted against the potential energy, rather than against $T_{\rm p}$. To this aim, we define a reference potential-energy-per-particle scale based on the crossover temperature $\Tx$, as $u(\Tx)$. We then plot in Fig.~\ref{fig:chi_utx} $\chi$ vs.~the dimensionless ratio $(u(T_{\rm p})\!-\!u(\Tx))/\Tx$ (in units such that the Boltzmann constant is unity); the saturation of $\chi$ at low energies is apparent in this representation, which allows comparison of the behavior of the different models on the same footing.  Notably, $\chi$ of the polydisperse soft spheres (pentagonal symbols in Fig.~\ref{fig:chi_utx}), which varies substantially over the sampled $T_{\rm p}$ range, shows a clear saturation at low energies, while featuring the same putative bound (as in Fig.~\ref{fig:chi_tptx}a, it is marked by the dashed horizontal line).

\vspace{1cm}
The simplicity and experimental relevance of $\chi$ as a dimensionless measure of mechanical noise makes it a highly relevant physical quantity to consider. Yet, it would be interesting and insightful to consider other quantifiers of mechanical disorder, testing their relation to $\chi$ and in particular elucidating whether they feature a lower bound as well. As mentioned in the introduction above, self-organized glasses generically feature soft, quasilocalized modes, whose frequency follows a universal nonphononic density of states ${\cal D}(\omega)\!=\!A_{\rm g}\omega^4$, see for example the glassy spectra shown in Fig.~\ref{fig:chi_ag}a for several values of $T_{\rm p}$. The prefactor $A_g$ physically represents the abundance of soft quasilocalized excitations in a glass, and has been shown~\cite{cge_paper,LB_modes_2019,pinching_pnas,MW_Ag_pre_2020} to vary over a huge range between low- and high-$T_{\rm p}$ computer glasses. As such, it strongly affects the mechanical properties of a glass, for example its plastic deformability and resistance to failure~\cite{itamar_brittle_to_ductile_pre_2011,Ozawa6656,loading_geometry_mrs_2021}.

In Fig.~\ref{fig:chi_ag}b, we plot $\chi$ against $A_{\rm g}$ for all studied glass models quenched from a melt, made dimensionless by multiplying it by $\omega_0^5\!\equiv\!(c_s/a_0)^5$, where $c_s$ is the shear wave speed and $a_0\!\equiv\!(V/N)^{1/3}$ is a characteristic interparticle distance. We find that plotting $\chi$ vs.~$A_{\rm g}$ reveals an approximate master curve, indicating a close relation between the two dimensionless quantifiers of mechanical disorder~\cite{loading_geometry_mrs_2021,phonon_widths2_pre_2021,ultrahigh_poissons_ratio_prm2022}. The apparent master curve shows a clear indication of saturation of $\chi$ towards lower $A_{\rm g}$'s, constituting additional supporting evidence for that $\chi$ satisfies a lower bound for glasses quenched from a melt.



\section{Summary and outlook}\label{sec:results_and_discussion}

In this work, we explored the variability of mechanical disorder quantified by a broadly applicable quantifier $\chi$ --- defined via the relative realization-to-realization fluctuations of the shear modulus ---, among different classes of disordered solids, ranging from systems in which disorder can be directly tuned, to systems in which disorder is an entirely emergent, self-organized property. The data sets generated for this work (concisely summarized in Table~\ref{tab:bigtable}) are envisioned to serve as references for future investigations, owing to the generality of the disorder quantifier $\chi$ which is applicable to any disordered solid, and allows to compare different model systems on the same footing. Importantly, $\chi$ is experimentally accessible through wave scattering measurements, as highlighted above. 

We explicitly demonstrated that mesoscopic mechanical disorder as captured by the quantifier $\chi$ is unbounded from above. In particular, as shown in Sect.~\ref{sec:networks}, $\chi$ is seen to diverge as the unjamming transition is approached. On the other hand, we presented evidence that in structural glasses quenched from a melt --- in which disorder is self-organized --- $\chi$ appears to be bounded from below in glasses quenched from very deeply supercooled states. The suggested saturation of $\chi$ at deep supercooling (low $T_{\rm p}$) implies that attenuation rates of long-wavelength acoustic excitations are bounded in the harmonic regime~\cite{scattering_jcp_2021}, when considered in terms of the typical elastic timescale of a glass. In addition, we expect related bounds on heat transport coefficients. 

Another interesting direction for future investigation is the effect of mesocopic mechanical disorder, e.g.~as quantified by $\chi$, on the elastic response of disordered solids for $r\!<\!\xi$, i.e.~on spatial scales before continuum elasticity is recovered. This basic issue seems to be particularly important for systems that undergo an unjamming transition, where $\xi$ can be large.   

\section*{Acknowledgements}
\noindent We thank Geert Kapteijns, David Richard, Corrado Rainone, Talya Vaknin, Avraham Moriel, and Gustavo D\"uring, for their comments on the manuscript. We greatly benefited from discussions with Gustavo D\"uring, David Richard, and Massimo Pica Ciamarra, who are warmly acknowledged. E.~B.~acknowledges support from the Ben May Center for Chemical Theory and Computation and the Harold Perlman Family. E.~L.~acknowledges support from the NWO (Vidi grant no.~680-47-554/3259).

\appendix

\section{Disordered crystals}
\label{sec:appendix_disordered_crystals}

We employ a simple model of disordered crystals in which soft, point particles interacting via a $\sim\! r^{-10}$ pairwise potential of the form
\begin{equation}
    \varphi(r,\lambda) = \left(\frac{\lambda}{r}\right)^{10} + \varphi_{\mbox{\tiny smooth}}(r/\lambda)\,.
\end{equation}
where $\varphi_{\mbox{\tiny smooth}}$ ensures smoothness and continuity at a cutoff interaction (see further details in Ref.~\cite{cge_paper}), and $\lambda$ is a microscopic interaction length. We place $N$ particles on a FCC lattice, randomly chose half of them that will be `inflated' by tuning the length parameters $\lambda_{\rm ls}$ and $\lambda_{\rm ll}$ of the inter- and inner-species interaction potential, respectively, to depend on the dimensionless parameter $\delta$ as
\begin{eqnarray}
    \lambda_{\rm ls} & = & \lambda_0(1\!+\!\delta/5)\,, \nonumber \\
    \lambda_{\rm ll} & = & \lambda_0(1\!+\!2\delta/5)\,, \nonumber 
\end{eqnarray}
and $\lambda_{\rm ss}\!=\!\lambda_0$, independent of $\delta$. With this construction, $\delta\!=\!0$ is a standard monodisperse crystal, and $\delta\!=\!1$ corresponds to a 1:1.4 binary mixture commonly used in glass-physics studies. Simulations are performed by incrementing $\delta$ by small increments, followed by a potential-energy minimization.

\section{Computer glass models}
\label{sec:appendix_models}


In this Appendix we provide brief descriptions of the models from sections~\ref{sec:fsp}-\ref{sec:melts} and refer to relevant literature where further information about the models and protocols used to prepare our ensembles of athermal glasses can be found. Number of particles $N$, number of independent glass samples $n$, and Poisson's ratio $\nu$ of systems in Sec.~\ref{sec:melts} are reported in Table~\ref{tab:poisson_and_sys_size}.\\

\noindent\textbf{Fluctuating-size-particles}.--- The `Fluctuating-size-particles' model allows generating computer glasses that are not formed by the conventional route of quenching a liquid. In this model, on top of the three translational degrees of freedom per particle, particles' effective sizes are allowed to fluctuate at some energetic cost. The latter is controlled by a stiffness $k_\lambda$; as $k_\lambda\!\rightarrow\!\infty$, the particles' sizes are frozen, and the model reduces to a conventional computer glass model. We use the glasses generated in~\cite{fsp}, where it was shown that $k_\lambda$ plays a crucial role in obtaining ultra-stable glasses by systematically varying $k_\lambda$ between $10^1$ to $10^5$. The number densities we employed were: $N/V\!=\!1.2660,0.8945,0.6765$,
$0.5801,0.5332,0.5166,0.5102,0.5082,0.5075$, expressed in terms of $\lambdabar^{-3}$ (see~\cite{fsp} for definition), for $k_\lambda\!=\!10$, 30, 100, 300, 1000, 3000, 10000, 30000, 100000, 300000, respectively.
Samples were quenched using the FIRE minimization algorithm~\cite{fire}.\\

\noindent{\textbf{Sticky hard spheres}}.--- Glasses made with this model were recently shown to have interesting elasticity properties~\cite{ultrahigh_poissons_ratio_prm2022}. Particles interact pairwise via a piece-wise potential constructed such that its repulsive part can be systematically controlled by the parameter $Q$ while keeping the attractions unchanged. The pairwise-potential reads:
\begin{equation}
    \frac{\varphi(r_{ij})}{\varepsilon}= 
        \begin{cases}
         \left(\frac{\lambda_{ij}}{r_{ij}} \right)^{Q} - \left(\frac{\lambda_{ij}}{r_{ij}} \right)^{6} + \bar{\varphi}_{\rm{\tiny{smooth}}}\left(r_{ij},Q\right) & \text{if } \frac{r_{ij}}{\lambda_{ij}}<2\\
        0 & \text{if } \frac{r_{ij}}{\lambda_{ij}}\geq 2
    \end{cases},
    \label{eq:hss_potential}
\end{equation}

where $\epsilon$ represents a microscopic energy scale, $\lambda_{ij}\!=\!\lambda, 1.18\lambda$, or $1.4\lambda$ for ‘small’-‘small’, ‘small-large’ or ‘large-
large’ interactions, respectively, and $\lambda$ forms the microscopic units of length. The function $\bar{\varphi}_{\rm{\tiny{smooth}}}\left(r_{ij},Q\right)\!=\!\sum_{\ell\!=\!0}^{2} c_{2\ell}\left(\frac{r_{ij}}{\lambda_{ij}}\right)^{2\ell}$ ensures that the potential and two derivatives
vanish continuously at the dimensionless cutoff distance
$r_{ij}\!/\!\lambda_{ij}\!=\!2$. For each $Q$ we used an ensemble of $1000$ glasses with $N\!=\!10976$ particles at a pressure-to-bulk modulus ratio of $p/K\!\approx\!10^{-3}$ or smaller. For more details on the molecular dynamics simulation see~\cite{ultrahigh_poissons_ratio_prm2022}. In table~\ref{tab:hs_coeffs_and_rhos} we provide the coefficients $c_{2\ell}$ and the dimensionless densities $\lambda^{3}N/V$ employed for the chosen $Q$s.\\

\begin{table*}
\centering
\caption{\label{tab:hs_coeffs_and_rhos}
\footnotesize Coefficients $c_{2\ell}$ of $\bar{\varphi}_{\tiny{\rm{smooth}}}\left(r_{ij},Q\right)$used to ensure the sticky hard spheres potential given by equation~\ref{eq:hss_potential} and two derivatives vanish continuously at the dimensionless cutoff for all $Q$ parameters studied in this work.}
\vspace{0.2cm}
{\renewcommand{\arraystretch}{1.2}
\begin{tabular}{|c |c |c |c |c |}
\hline
$Q$ & $c_{4}$ & $c_{2}$ & $c_{0}$ & $\lambda^{3}N\!/\!V$\\
\hline
10 & 0.0189208984375 & -0.20263671875 & 0.611328125 & 0.777\\
\hline
20 & 0.014647617936134338 & -0.15624284744262695 & 0.4687342643737793 & 0.737\\
\hline
40 & 0.013020833332006987 & -0.13888888887777284 & 0.41666666664332297 & 0.708\\
\hline
80 &  0.012335526315789474 & -0.13157894736842105 & 0.39473684210526316 & 0.694\\
\hline
160 & 0.01201923076923077 & -0.1282051282051282 & 0.38461538461538464 & 0.6802\\
\hline
320 & 0.011867088607594937 & -0.12658227848101267 & 0.379746835443038 & 0.677\\
\hline
\end{tabular}
}
\end{table*}

\noindent{\textbf{Sticky spheres (binary)}}.--- This is a $50\!:\!50$ binary mixture of `small' and `large' particles that interact pairwise via a piece-wise Lennard-Jones-like potential, introduced in first in~\cite{potential_itamar_pre_2011}, and studied extensively in~\cite{itamar_brittle_to_ductile_pre_2011,Massimo_supercooled_PRL,sticky_spheres_part_1,sticky_spheres_part2}. In this model, the strength of attractions between the glass's constituent particles can be readily tuned by varying the interaction cutoff-length $r_{\rm c}$. In this work, we employ three different cutoff-lengths: $r_{\rm c}\!=\!1.2,1.3,1.5$ enumerated in Figs.~\ref{fig:chi_tptx} and ~\ref{fig:chi_utx} of the main text as '1,2,3', respectively. The number density is fixed to $N/V\!=\!0.60\lambdabar^{-3}$ for all of the `1,2,3' ($r_{\rm c}\!=\!1.2,1.3,1.5$) variants of the model, where $\lambdabar$ denotes the effective size of the `small' species. For further information about the model and glass preparation protocol see~\cite{sticky_spheres_part2}.\\

\noindent{\textbf{Sticky spheres (polydisperse)}}.---In this model, pairs of particles interact via the same pairwise potential as in the `Sticky spheres (binary)' model discussed above, in which the cutoff length $r_{\rm c}$ serves as the key control parameter. We follow~\cite{LB_swap_prx} and draw each particle's effective size $\lambda_i$ from a distribution $p(\lambda)\!\sim\!\lambda^{-3}$, between $\lambda_{\rm min}\!=\!1.0\lambdabar$ and $\lambda_{\rm max}\!=\!2.22\lambdabar$, where $\lambdabar$ is the simulation units of length. We study two variants of this model that differ in their respective cutoff-lengths: $r_{\rm c}\!=\!1.1$, referred to as `sticky spheres (polydisperse) 1' in the main text, and $r_{\rm c}\!=\!1.2$, referred to as 'sticky spheres (polydisperse) 2' in the main text. The models' polydispersity is used to exploit the power of the Swap Monte Carlo Method~\cite{LB_swap_prx} to obtain the deeply supercooled equilibrium states. Finite-size effects induced by random particle-sizes are dealt with as explained in~\cite{boring_paper}. We choose the number density $N/V\!=\!0.40\lambdabar^{-3}$ such that the high parent-temperature glasses' pressure to bulk modulus ratio $p/K\!\approx\!0.05$.\\

\noindent{\textbf{Polydisperse soft spheres}}.--- This model consists of particles of random effective sizes that are distributed and chosen in the same way as done for the Sticky spheres (polydisperse) model described above. Pairs of particles interact via an inverse-power-law (IPL) pairwise potential $\varphi_{\mbox{\tiny IPL}}\!\sim\!r^{-10}$ ($r$ is the interparticle separation), truncated and smoothed as explained in e.g.~\cite{boring_paper}. Equilibrium states were obtained using the Swap Monte Carlo Method~\cite{LB_swap_prx}, which allows one to obtain very deeply supercooled states. The number density was fixed at $N/V\!=\!0.58\lambdabar^{-3}$ for all systems. This model was studied extensively in~\cite{boring_paper,pinching_pnas} where further details can be found, including an explanation about how the polydisperisty is chosen, and how disorder-realization-induced finite-size effects are handled.\\

\noindent{\textbf{Hertzian elastic spheres}}.--- The Hertzian elastic spheres model is a $50\!:\!50$ binary mixture of small and large soft, linear-elastic spheres interacting via the Hertzian interaction law~\cite{landau1964theory}. The simulations are performed
at a fixed number density $N/V\!=\!0.9386\lambdabar^{-3}$, where $\lambdabar$ denotes the diameter of the small particles. For this number density, and for glasses of high $T_{\rm p}$'s, the typical ratio of the pressure to bulk modulus is $p/K\!\approx\!0.17$. Information about the parameters employed and further details can be found in~\cite{modes_prl_2020}.\\

\noindent{\textbf{Kob-Anderesen binary Lennard-Jones}}.---This model is probably the most thouroughly studied computer glass-former~\cite{kablj2_mct_test}. This canonical glass-former is a binary mixture of $80\%$ type A (`large') particles, to $20\%$ type B (`small') particles, which interact via the standard 12-6 Lennard-Jones potential. The potential was truncated smoothly up to the first derivative~\cite{sri_nat_comm_2017}.\\

\begin{table}[h!]
\centering
\caption{\label{tab:poisson_and_sys_size}
\footnotesize High $T_{\rm{p}}$ Poisson's ratio $\nu$ sorted in descending order, system size $N$, and number of independent samples $n$ used to estimate $\chi$, for the 9 computer glass-forming models employed in this work. The difference between the employed models can be appreciated from the variability of $\nu$, which shows a difference up to $\approx 28\%$ among models. For the Fluctuating-size-particles model with $k_{\lambda}\!=\!30$, $n\!=\!10000$ (see Appendix~\ref{sec:appendix_models} for further details).}
\vspace{0.2cm}
{\renewcommand{\arraystretch}{1.2}
\begin{tabular}{|c|c|c|c|}
\hline
computer glass model & $\nu$ & $N$ & $n$\\
\hline
Polydisperse soft spheres & $0.4293$ & $16000$ & $2000$\\
\hline
Sticky spheres (binary) 3 & $0.4034$ & $10000$ & $3000$\\
\hline
Hertzian elastic spheres & $0.4015$ & $4000$ & $10000$ \\
\hline
Fluctuating-size-particles & $0.3952$ & $4000$ & $42000$\\
\hline
Kob-Andersen binary Lennard-Jones & $0.3888$ & $3000$ & $8970$\\
\hline
Sticky spheres (binary) 2 & $0.3782$ & $3000$ & $9200$ \\
\hline
Sticky spheres (binary) 1  & $0.3436$ & $3000$ & $9200$ \\
\hline
Sticky spheres (polydisperse) 2 & $0.3229$ & $2000$ & $11100$\\
\hline
Sticky spheres (polydisperse) 1 & $0.3061$ & $2000$ & $8900$\\
\hline
\end{tabular}
}
\end{table}
\vspace{2cm}
\section{Finite size effects in $\chi$}
\label{sec:appendix_finitesize_chi}
\begin{figure}[h!]
	\centering
	\includegraphics[width=0.9\linewidth]{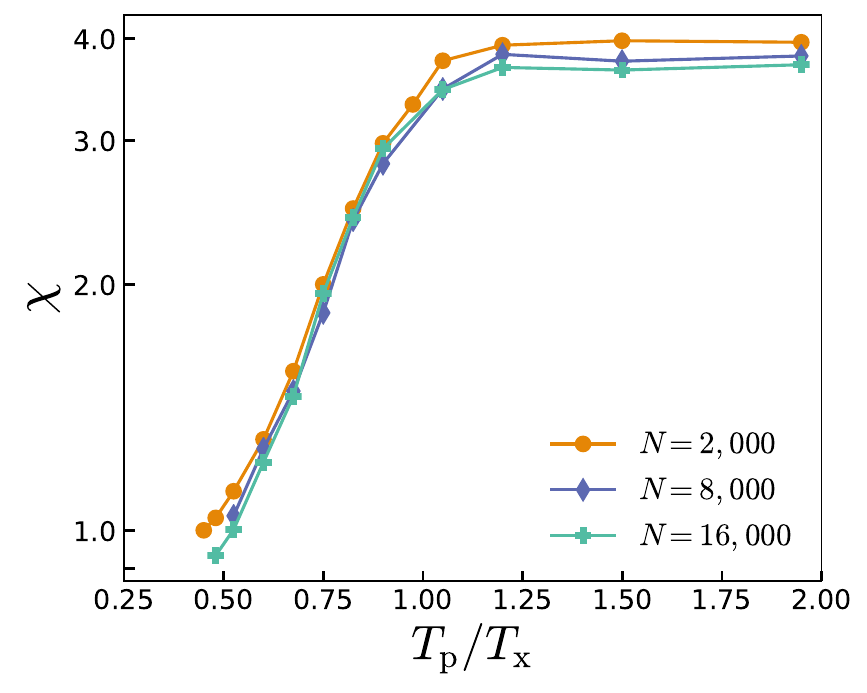}
	\caption{\small{Finite size effects are present in $\chi$.} Here we show the dimensionless quantifier $\chi$ estimated following the scheme put forward in~\cite{sticky_spheres_part_1} --- also explained in the main text (see Sect.~\ref{sec:appendix_models}) --- and plot it against the rescaled parent temperature $T_{\rm p}/\Tx$ for the Polydisperse soft spheres model (notice the semi-log scale).}
	\label{fig:finite-size}
\end{figure}

Since the probability distribution function of the shear modulus $G$ is known to feature strong finite-size effects~\cite{scattering_jcp_2021}, the disorder quantifier $\chi$ was estimated using the Jackknife-like method described in Sec.~\ref{sec:chi_calculation} and in Ref.~\cite{sticky_spheres_part_1}. Fig.~\ref{fig:finite-size} shows the presence of such finite-size effects in $\chi$, which seem to be independent of the parent temperature $T_{\rm p}$, contrary to what one would expect due to the associated decrease of the length scale $\xi$~\cite{pinching_pnas,finite_size_modes_pre_2020}. 

\section{Crossover temperature $\Tx$ and interpolation of the potential energy per particle $u(\Tx)$}
\label{sec:appendix_interpolation_energy}
To compare the $T_{\rm p}$ dependence of the statistical mechanical properties of our diverse computer glasses from Sec.~\ref{sec:melts} on the same footing, a characteristic temperature scale for each glass is needed. Since $\chi$ quantifies mechanical disorder in terms of the shear modulus fluctuations, such a temperature scale should be relevant to the elastic properties of glasses. Here, we employ a recently introduced crossover temperature scale $\Tx$, which has been shown to control the $T_{\rm p}$ dependence of elastic properties in computer glasses (see~\cite{Tx_jcp}).

To extract $\Tx$, we follow~\cite{Tx_jcp}: for a given computer liquid, we consider a large ensemble of equilibrium liquid states at temperature $T$. For each such state, we follow steepest-descent dynamics and record the mean and standard deviation of particles' squared displacements $\overline{\delta r^2}$ between the initial, equilibrium state, to the inherent state that is inevitably reached~\cite{Sastry1998}. The ratio of the aforementioned standard deviation and mean of the squared displacements forms a dimensionless number, which is a nonmonotonic function of the equilibrium temperature $T$~\cite{Tx_jcp}. The temperature at which this function assumes a maximum is defined as the crossover temperature, with respect to which all temperatures in our work are normalized.

\begin{figure}
	\includegraphics[width=1.0\linewidth]{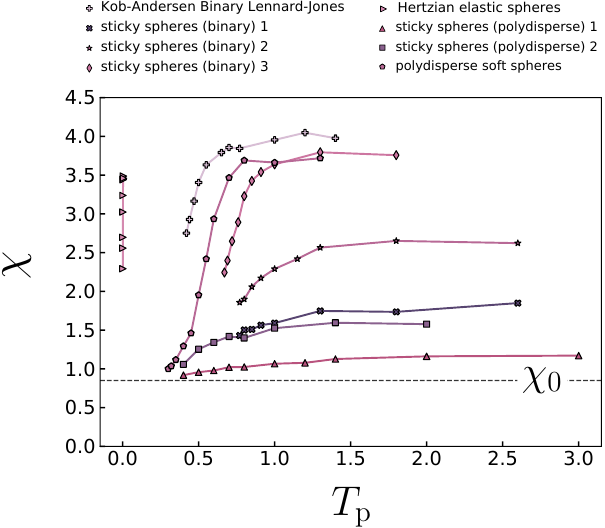}
	\caption{\small The same data of Fig.~\ref{fig:chi_tptx}a of the main text are plotted against the computer $T_{\rm p}$ expressed in each model's microscopic/simulational units. This representation shows that the independently measured crossover temperature $\Tx$ is effective in organizing the elastic properties of model glasses, as also demonstrated in~\cite{Tx_jcp}.}
	\label{fig:chi_vs_raw_tp}
\end{figure}

The estimated values of $\Tx$ --- expressed in terms of each model's individual simulation units --- are reported in Appendix~\ref{sec:appendix_interpolation_energy} (see Table~\ref{tab:Tx_utx}). Additionally, we corroborate therein the usefulness of $\Tx$ in putting the elastic properties of model glasses on the same footing (see Fig.~\ref{fig:chi_vs_raw_tp}).

While there are temperature scales extracted from and relevant to supercooled liquids’ dynamics, here we employed a temperature scale $\Tx$ that has been shown to be relevant to the elastic properties of the glassy states~\cite{Tx_jcp}. In Fig.~\ref{fig:chi_vs_raw_tp}, we show $\chi$ as a function of {\it{only}} the parent temperature $T_{\rm{p}}$, which supports the utility of the crossover temperature $\Tx$ in organizing
elasticity data. Table~\ref{tab:Tx_utx} reports the crossover temperature with respect to which all temperatures in our work are normalized.

In Fig.~\ref{fig:chi_utx} of the main text, we plot $\chi$ vs.~$\big(u(T_{\rm p}) - u(\Tx)\big)/\Tx$, where the energy per particle at the crossover temperature $u(\Tx)$ was obtained via a linear interpolation of the form
\begin{equation}
    u(\Tx) = u(T_{1}) + \frac{\Delta u}{\Delta T} (\Tx-T_{1});
\end{equation}
where $\Delta u \!=\! u(T_{2})\!-\!u(T_{1})$, $\Delta T\!=\!T_{2}\!-\!T_{1}$, and $T_1\!\le\!\Tx\!\le\!T_2$. The extrapolated values for the 8 models are reported in Table~\ref{tab:Tx_utx}.

\begin{table}[h!]
\centering
\caption{\label{tab:Tx_utx}
\footnotesize Crossover temperatures $\Tx$ as estimated for the computer glass models employed in this work, following the approach presented in~\cite{Tx_jcp} and interpolated values for the energy per particle at the crossover temperature $u(\Tx)$ for our different models of computer glasses quenched from equilibrium states. Both quantities are expressed in \emph{simulational units}.}
\vspace{0.1cm}
{\renewcommand{\arraystretch}{1.2}
\begin{tabular}{|c|c|c|}
\hline
computer glass model & $\Tx$ & $u(\Tx)$\\
\hline
Sticky spheres (binary) 1   & $1.26$ & $-4.583$ \\
\hline
Sticky spheres (binary) 2 & $1.06$ & $-4.884$ \\
\hline
Sticky spheres (binary) 3 & $0.85$ & $-5.224$\\
\hline
Sticky spheres (polydisperse) 1 & $1.27$ & $-4.233$\\
\hline
Sticky spheres (polydisperse) 2 & $0.921$ & $-4.876$\\
\hline
Polydisperse soft spheres & $0.66$ & $5.262$\\ 
\hline
Hertzian elastic spheres & $0.0023$ & $0.008$ \\
\hline
Kob-Andersen binary Lennard-Jones & $0.535$ & $-6.953$\\
\hline
\end{tabular}
}
\end{table}



%
\end{document}